\documentclass[pra,twocolumn,amsmath,amssymb]{revtex4}
\usepackage{bm}

\usepackage{epsfig}

\begin{document}

\title{Effective Hamiltonians in quantum physics: resonances and geometric phase}

\author{A.\  R.\  P.\  Rau${}^{*}$ and D. Uskov \\
Department of Physics and Astronomy, 
Louisiana State University, 
Baton Rouge, Louisiana 70803-4001}

\date{\today}

\begin{abstract}

Effective Hamiltonians are often used in quantum physics, both in time dependent and time independent contexts. Analogies are drawn between the two usages, the discussion framed particularly for the geometric phase of a time-dependent Hamiltonian and for resonances as stationary states of a time-independent Hamiltonian. 
\end{abstract}

\maketitle
    
\section{Introduction}

Effective Hamiltonians can arise in a variety of contexts. Chosen to focus on some particular aspect or sub-system, an effective Hamiltonian $H_{\rm eff}$ is constructed from the full Hamiltonian $H$ of a physical system. Often, $H_{\rm eff}$ is simpler and of dimension smaller than $H$. The use of complex energies, with the imaginary part a stand-in for the usually complicated and infinite-dimensional aspects of friction or dissipation, is a familiar example, occurring in various areas of physics${}^{1}$. Both the time-dependent and the time-independent Schr\"{o}dinger equation admit descriptions in terms of effective Hamiltonians, forming the theme of this article.

\section{Time independent case}

Consider first a time-independent situation, namely, stationary states of a time-independent Hamiltonian $H$. These may include discrete bound states with negative energy eigenvalues and scattering states at positive energy but also resonances in that positive energy sector, which are ``quasi-bound" states${}^2$. Resonances are typically viewed in terms of a decomposition $H=H_0 +V$, and the basis states provided by the eigenfunctions of $H_0$. These will include both bound and continuum states. The ``interaction" $V$, which is contained within the full $H$, mixes these states of $H_0$, such a superposition constituting the quasi-bound resonance state of the full $H$. Because of this superposition, the resonance state has both discrete and continuum character. Further, the inclusion of the continuum means that the superposition necessarily embraces an infinity of states.

The splitting $H=H_0 +V$ may be either in real configuration space or in state space. The former can arise, for instance, when the potential in a two-particle system goes asymptotically to zero so that for any positive energy the particles separate to infinity, but intervening barriers may temporarily trap the system before quantum tunneling allows escape. This last phrase with its ``before" and ``temporarily" introduces a time aspect into a time-independent problem. While unnecessary, this exemplifies the role of complementary pictures in quantum physics. The problem as a whole is of time-independent stationary states, some of which are characterized not just by an energy position but also a width (tunneling or otherwise) and, possibly, other real parameters as well (the so-called ``profile index" $q$ of an asymmetric resonance being an example${}^{3,4}$). The energy and width may be subsumed into a complex energy $E_r +i\Gamma /2$, although there is nothing intrinsically complex about the problem of stationary states of a real $H$. The width $\Gamma$ in a time-independent picture is complementarily related to the lifetime of the resonance in terms of a time evolution. Examples of potentials with intervening barriers are many: alpha-decay of nuclei, electric field ionization, two-valley potentials with angular momentum barriers in atoms and molecules${}^3$, etc. See Fig. 1. In any of these systems, if the radial variable $r$ (or a hyperspherical equivalent in a many-particle system${}^3$) is split into an inner and an outer region so that the former includes the barriers, even the states of positive $E_r$ are bound states within that region. But they leak out into the outer region in the full problem and are, indeed resonances. Because of the key role played by shapes of potentials that give rise to them, they are called ``shape" resonances.

\begin{figure}
\scalebox{2.0}{\includegraphics[width=1.7in]{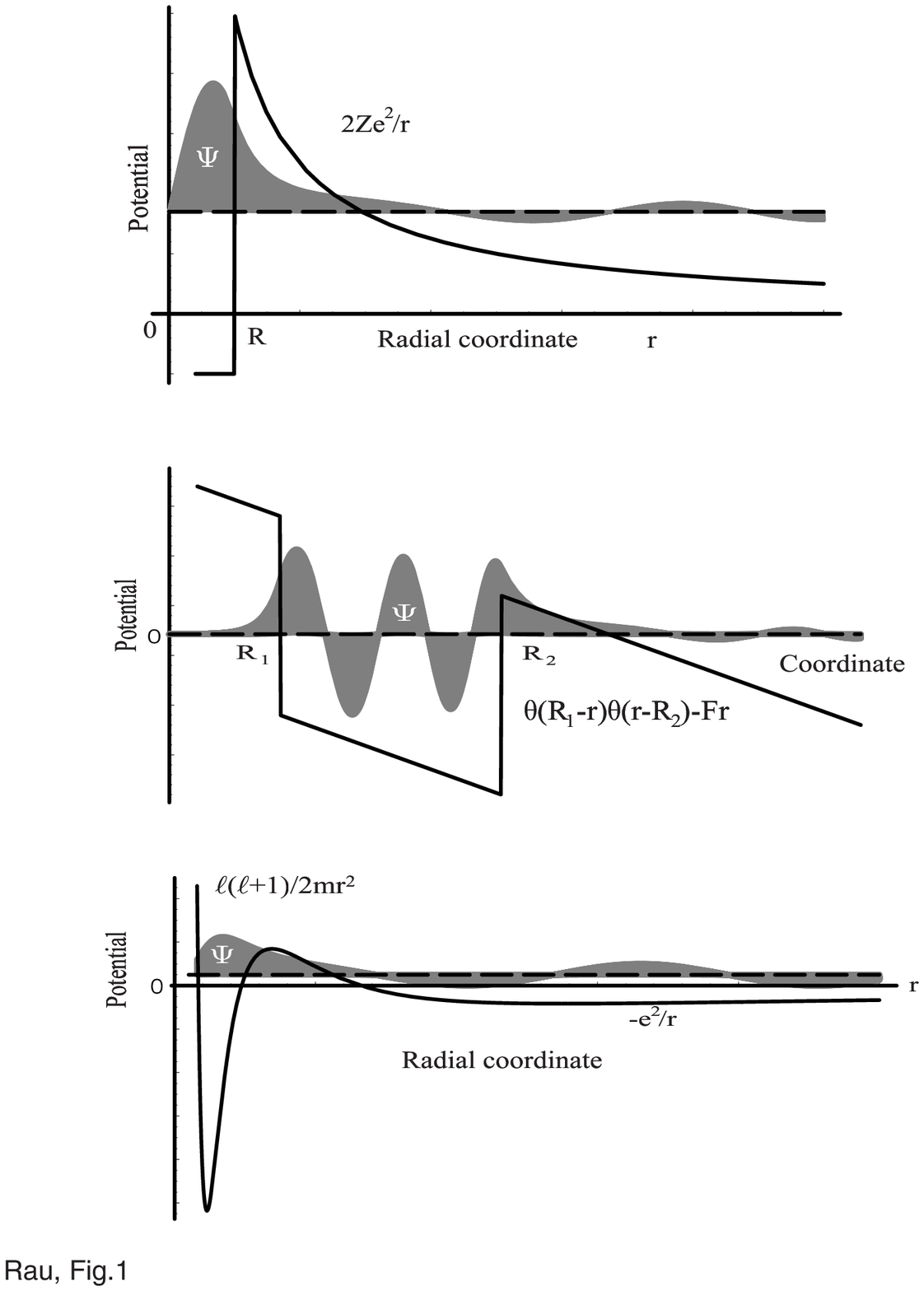}}
\caption{Potentials with intervening barriers leading to shape resonances with their quasi-bound wave functions shown. (a) alpha-decay of nucleus of charge $(Z+2)$ with schematic inner nuclear attraction and Coulomb repulsion of separating particles. (b) Field ionization from metals in an electric field F of states near the Fermi energy shown as zero. (c) Two valley potentials in atoms and molecules formed from superposition of angular momentum and atomic or molecular potentials, typically for $\ell \geq 2$.}
\end{figure}

Instead of the splitting in the above paragraph into two parts in $r$, the $H=H_0 +V$ breakdown may be in terms of states. Thus, consider doubly-excited states in atoms and molecules, the first of these in the helium atom serving as a good example${}^3$. As in Fig. 2, consider states of this two-electron system at energies around 58 eV above the ground state of the atom. In $^{1}S^e$ symmetry, these being good quantum numbers of the whole system (we will consider the non-relativistic $H$, ignoring spin-orbit or other relativistic complications), there lie here the states of the one-electron ionization continuum built on the He$^{+} 1s$ ionic ground state. In an independent electron description, that is, when $H_0$ includes only one-electron terms and possibly a mean field of the electron-electron interaction, with the residual part of this interaction constituting $V$, such states can be designated $1sks \,^{1}S^e$, where $k$ is the wave-number of the continuum electron related to its kinetic energy through $E=(\hbar k)^2/2m$. The values of $E$ are, approximately, (58-24.6 = 33.4 eV). But there also lie in this region bound states of $H_0$ that may be described as $2s^2 \, ^{1}S^e$ and $2p^2 \, ^{1}S^e$, which are states of an electron bound to He$^{+} 2s$ or $2p$ that share the same overall symmetry. Thus, in terms of such an independent electron description, all these states are degenerate and, by virtue of the residual $V$, are superposed in the physical eigenstates of the full $H$. These are the doubly-excited ``Feshbach" resonances${}^5$, having both discrete and continuum character in their description, that may be seen either in excitation cross-sections from the ground state (with 58 eV of excitation energy delivered by some means) or in elastic scattering of electrons from He$^{+} 1s$ around 33.4 eV.

\begin{figure}[h]
\scalebox{2.0}{\includegraphics[width=2in]{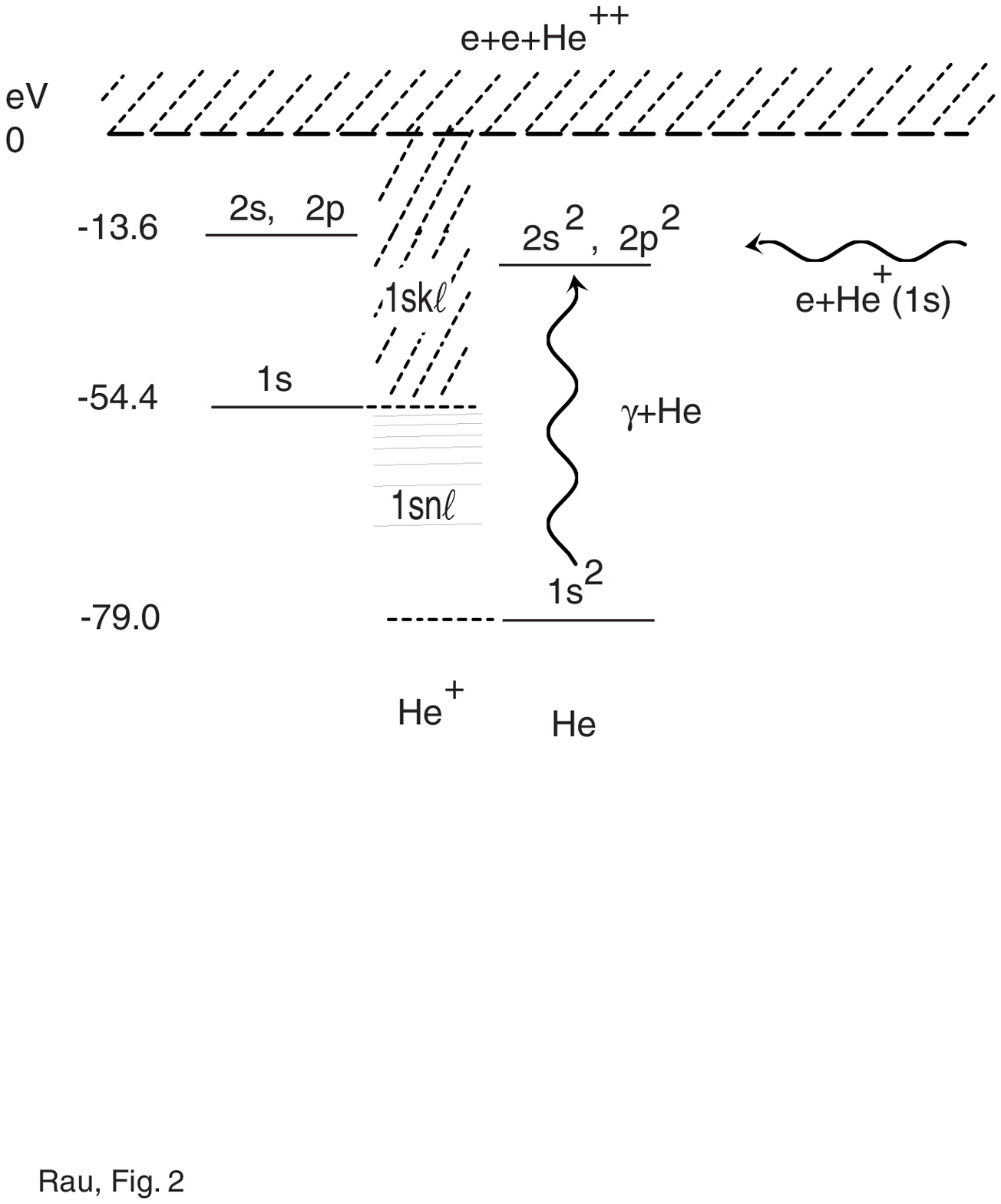}}
\caption{Ground and first excited state (there are infinitely many) of He$^+$ and the lowest bound states of He (again there are infinitely many) attached to them are shown, along with their energies below the fully dissociated limit of two electrons and the helium nucleus. Cross-hatched region is the ionization continuum of He built on the ground state of He$^+$, that is, the two-electron states $1sks$, $E=\frac{1}{2}k^2$ being the energy of the continuum electron. The $2s^2$ and $2p^2$ states lie embedded in this continuum, that is, are degenerate with $1sks$ and, therefore, mix with it to give the quasi-bound resonance states. The resonances can be accessed either by (e + He$^+$) scattering or by photoexcitation from the ground state as shown (because of dipole selection rules, a single photon will reach similar states of $2s2p \,^{1}P^o$ symmetry whereas two-photon absorption would be necessary to reach the $2s^2/2p^2 \,{}^1 S$ states).}
\end{figure} 

In summary of the above two paragraphs, the two alternative pictures are but just that, our pictures or descriptions. The physical system is one of eigenstates of the three-body Hamiltonian $H$ in a certain energy range and is an integral whole. However, in our handling of them, and even more in the pictures we develop for our understanding, we develop alternative breakdowns in terms of simpler sub-systems, either in the real configuration space of $r$ or in two-electron states of an independent particle picture. Even as we do so, pictures and models being necessarily involved in our approach to any physical situation, we must also keep in mind that $H_0$, $V$, state configurations $2s^2$, $2p^2$ and $1sks$, etc., are not elements of the underlying physical reality nor are they accessible to measurement. In the same vein, the distinction between the terms ``shape" and ``Feshbach" is also somewhat arbitrary but, nevertheless, useful. In particular, with reference to Figs. 1 and 2, whereas the former often lie just above the energy threshold to which they are attached, the latter lie just below (the He$^{+}2s$ threshold). Correspondingly, the shape resonances tend to be broad (shorter lived) whereas the Feshbach resonances are typically narrow (longer lived). In a closely-related two-electron system to our discussion above, namely the negative ion of the hydrogen atom, and in the symmetry $^{1}P^o$ which can be reached by one-photon absorption from the ground state (unlike $^{1}S^e$ which takes two photons), an example of each was clearly shown on either side of the H$(n=2)$ threshold by a classic experiment${}^6$. See Fig. 3.

\begin{figure}[h]
\scalebox{2.0}{\includegraphics[width=2in]{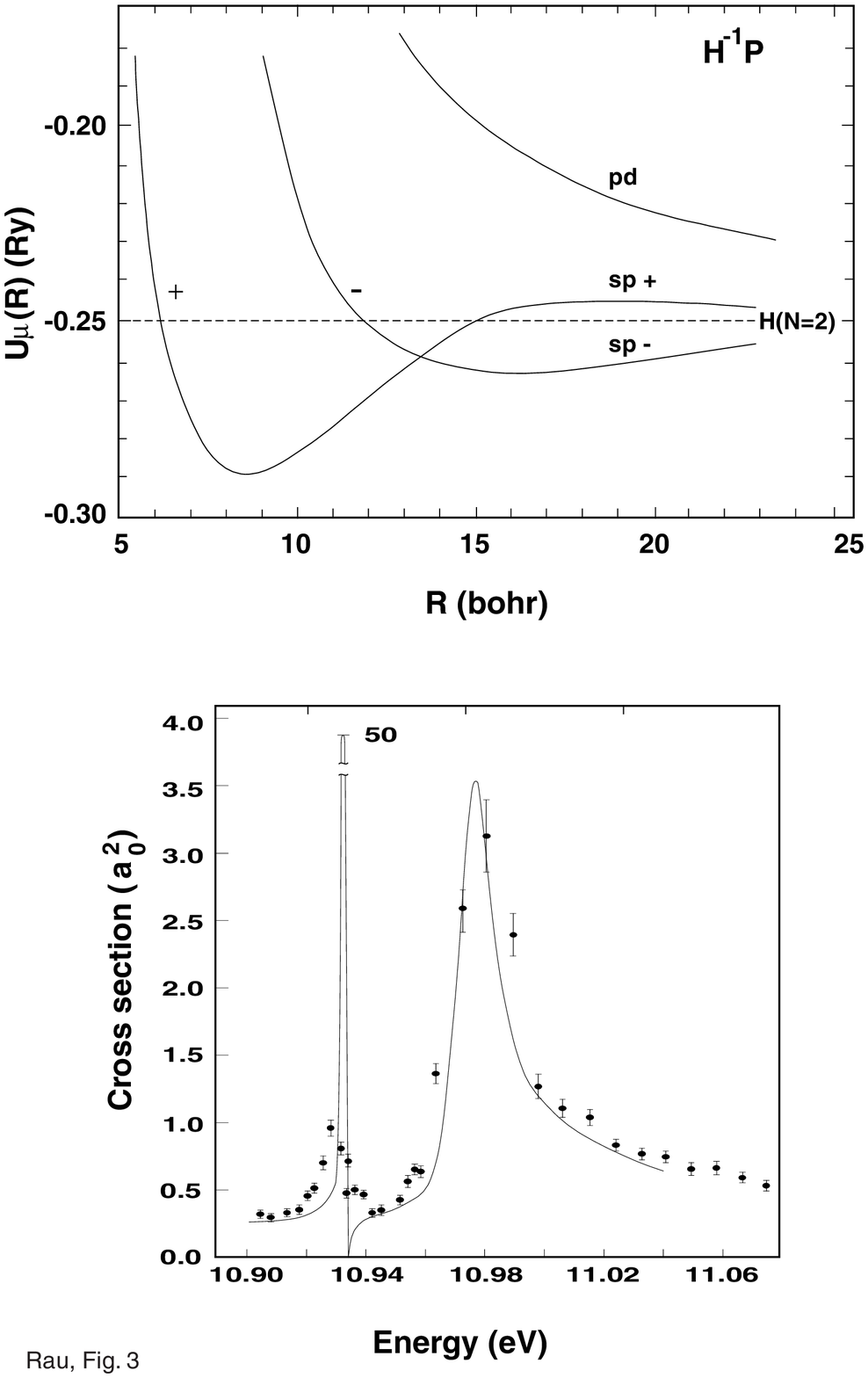}}
\caption{Theoretical hyperspherical potential wells and experimental observations of doubly-excited ${}^1P^o$ resonances of H${}^{-}$ in the vicinity of the H($N=2$) threshold as observed in  photodetachment. The cross-section (in units of squared Bohr radius) shows a sharp Feshbach resonance just below and a broad shape resonance just above the 10.95 eV threshold (from Ref.6). The potential wells (1 Ry =13.6 eV) in hyperspherical coordinates for the two-electron system account for the observed resonances, the barrier in the $+$ curve holding the shape resonance while the $-$ well supports the Feshbach resonance${}^7$.}
\end{figure}

The partitioning of a Hamiltonian into two spaces is conveniently done in the ``Feshbach projection operator" formalism${}^8$. With $P$ and $Q$ projection operators into what are usually referred to as ``closed" and ``open" subspaces (for alternative terminologies, see Section 8.6.1 and 8.6.2 of Ref. 3), the former a finite space with bound eigenstates and the latter the infinite space of the continuum into which decay takes place, we can rewrite $H\psi =E\psi$ as

\begin{equation}
H(P+Q)\psi = E(P+Q)\psi, \,\, P+Q =\mathcal{I}.
\label{eqn1}
\end{equation}
Multiplying from the left by $Q$ and re-arranging with the use of $Q^2=Q$, a projection property, we have 

\begin{equation}
Q\psi = (E-QHQ)^{-1}(QHP)\psi.
\label{eqn2}
\end{equation}
Multiplying Eq.~(\ref{eqn1}) by $P$ from the left, and substituting for $Q\psi$ from Eq.~(\ref{eqn2}) gives

\begin{equation}
[PHP +PHQ (E-QHQ)^{-1} QHP)]P \psi= EP \psi.
\label{eqn3}
\end{equation}
The term in square brackets is the effective Hamiltonian $H_{\rm eff}$ in $P$-space. The term ``optical potential" is also used. Note that it has $P$ at both right and left extremes. With $\psi$ also occurring only in its projected piece, the Eq.~(\ref{eqn3}) is entirely in $P$-space. With no approximation made, the second term in $H_{\rm eff}$ includes the effect of the remaining $Q$-space. Note its structure in the form of a second-order energy, with $H$ ``carrying" from $P$ to $Q$ space and, then with an attached energy denominator, {\it entirely in that $Q$-space}, the $H$ ``returning" again to $P$-space. The first $PHP$ term has a purely discrete spectrum but some of those eigenvalues may acquire both a shift and a width as a result of the second term, which involves coupling to the $Q$-space. These are the resonances. Because of the energy denominator, the effect of this second term can be dramatic at energy values $E$ close to an eigenvalue of $QHQ$. 

\section{Time dependent case}

Turn next to the time-dependent Schr\"{o}dinger equation, $i\dot{U}(t)=H(t)U(t)$, with a dot standing for a time derivative, and where we choose to work with the evolution operator $U$ rather than the wave function $\psi(t)$, to which it is easily related through $\psi(t) =U(t) \psi(0), U(0)=\mathcal{I}$. Upon writing 

\begin{equation}
U(t) =U_1(t) \, U_2 (t),
\label{eqn4}
\end{equation}
taking a derivative with respect to time, and multiplying from the left by $U_1^{-1}$, we get

\begin{equation}
i\dot{U}_2 = H_{eff} U_2, \,\,H_{eff}=U_1^{-1}HU_1 -iU_1^{-1}\dot{U}_1.
\label{eqn5}
\end{equation}
Again, we have a reduced equation for $U_2$ alone but, of course, the effective Hamiltonian incorporates the part contained in $U_1$ so that the reduced expression is formally complete with no approximation implied. A connection to variational principles and identities will be discussed at the end of this Comment.

The above is completely general for any time-dependent problem. One important specific application is in separating the geometric phase${}^9$ from other elements of time evolution. With the advent of fault-tolerant quantum computation, the geometric phase is seen to have advantages in robustness and fidelity over dynamical phases that accumulate as a result of energy changes in time${}^{10}$. Consider, as an illustration, an angular momentum that couples through its magnetic moment to a magnetic field $\vec{B}(t)$, with only linear coupling in $H=-\vec{J} \cdot \vec{B}(t)$. Given the  three operators in the system, conveniently chosen as the usual triad $(J_{+} \equiv J_x+iJ_y, J_{-} \equiv J_x-iJ_y, J_z)$, a complete solution for the evolution operator can be obtained by writing${}^{11}$

\begin{equation}
U(t) = e^{ -i\mu_3 (t) J_{+}} e^{- i\mu_{2}(t)J_{-}} e^{- i\mu_1(t)J_z}.
\label{eqn6}
\end{equation}
That this is indeed the solution can be seen by a construction that leads to the required equations that define the functions $\mu$ in the exponents. Taking the time derivative, repeatedly applying a standard, ``Baker-Campbell-Hausdorff" (BCH), identity${}^{12}$ for $e^{A}Be^{-A}$ to cast as an operator multiplying $U$ from the left, gives

\begin{eqnarray}
i\dot{U} & = & [\dot{\mu}_{3}J_{+} +(J_z+i\mu_{3}J_{+})\dot{\mu}_1  \nonumber \\
 & + & (\dot{\mu}_{2}-i\mu_{2}\dot{\mu}_1)(J_{-}-2i\mu_{3}J_z+{\mu_{3}}^2 J_{+})]U.
\label{eqn7}
\end{eqnarray}
Upon identifying the right-hand side of Eq.~(\ref{eqn7}) with $HU$, the equations satisfied by the three $\mu$ follow:

\begin{eqnarray}
\dot{\mu}_{3} +\frac{1}{2}{\mu_{3}}^2 B_{+} -i\mu_{3}B_3 & = & -\frac{1}{2}B_{-}, 
\nonumber \\
\dot{\mu}_{2} -i\mu_{2}\dot{\mu}_1 =-\frac{1}{2}B_{+},  &  & \dot{\mu_1} 
+iB_{+}\mu_{3} = -B_3,
\label{eqn8}
\end{eqnarray}
where $B_{\pm} \equiv B_x \pm iB_y$. The requirement $U(0)=\mathcal{I}$ sets the boundary conditions for all three functions, $\mu (0)=0$.

Thus, solutions of the set of classical equations in Eq.~(\ref{eqn8}), when inserted into Eq.~(\ref{eqn6}), give the full quantum evolution. The first equation for $\mu_{+}$ decouples from the other two and may be solved by itself first, followed by simple quadrature of the other two. Note that any value of spin-$j$ obeys the same set of equations in Eq.~(\ref{eqn8}), since no use was made of any specific representation of the operators $J$, only their commutators. Further, $U$ is guaranteed to be unitary by construction${}^{11}$. Indeed, this implies relationships between the three complex $\mu$: 

\begin{equation}
\mu_{2}=\mu_{3}^*/(1+|\mu_{3}|^2), \,\, e^{{\rm Im} \, \mu_1} =  (1+|\mu_{3}|^2),
\label{eqn9}
\end{equation}
so that there are only three linearly independent quantities which may be chosen as the real and imaginary parts of $\mu_{3}$ and Re $\mu_1$. With this, for $j=1/2$, Eq.~(\ref{eqn6}) may be written as 

\begin{equation}
U(t) = \frac{e^{-i\mu_1 /2}}{1+|\mu_{3}|^2} \left(
\begin{array}{cc}
1 & -i\mu_{3}e^{i{\rm Re}\, \mu_1} \\
-i\mu_{3}^{*} & e^{i{\rm Re}\, \mu_1}
\end{array}
\right).
\label{eqn10}
\end{equation} 
Thus, an initial state 
$\psi (0)= \left( 
\begin{array}{c}
1 \\ 
0
\end{array}
\right)$
evolves to
 
\begin{equation}
\psi (t)=  \frac{e^{-i\mu_1 /2}}{1+|\mu_{3}|^2} \left(
\begin{array}{c}
1 \\
-i\mu_{3}^{*}
\end{array}
\right),
\label{eqn11}
\end{equation}
and the density matrix $\rho(t) =|\psi (t)\rangle \langle \psi (t)|$ becomes

\begin{equation}
\rho (t) = \frac{1}{1+|\mu_{3}|^2} \left(
\begin{array}{cc}
1 & i\mu_{3} \\
-i\mu_{3}^{*} & |\mu_{3}|^2
\end{array}
\right).
\label{eqn12}
\end{equation}
Whereas the density matrix involves only the two parameters of the complex quantity $\mu_{3}$, and so does the wave function (except for a usually unobservable phase $\mu_1$), the evolution operator depends on the third parameter, Re $\mu_1$, as well. The above equations give the $2 \times 2$ matrices for a spin-1/2 but all these features of the role of the three parameters contained in $\mu$ apply also to the $(2j+1)$ vectors and matrices of any spin-$j$.

The phase $\mu_1$ in the above expressions, particularly in the evolution operator $U$ in Eq.~(\ref{eqn6}) may be viewed as an illustration of Eq.~(\ref{eqn4}), where $U_1$ is the product of the first two exponential terms in Eq.~(\ref{eqn6}) and $U_2$ the last term involving $\mu_1$ alone. We have, upon using Eq.~(\ref{eqn9}), 

\begin{equation}
U_1=\left(
\begin{array}{cc}
\frac{1}{1+|\mu_{3}|^2} & -i\mu_{3} \\
\frac{-i\mu_{3}^{*}}{1+|\mu_{3}|^2} & 1
\end{array}
\right),
\label{eqn13}
\end{equation}
which depends only on $\mu_{3}$ whereas $\mu_1$ is contained in 

\begin{equation}
U_2=\left(
\begin{array}{cc}
e^{-i\mu_1 /2} & 0 \\
0 & e^{i\mu_1 /2}
\end{array}
\right).
\label{eqn14}
\end{equation}
 
It is instructive to see the specific form, including of individual terms, that $H_{\rm eff}$ in Eq.~(\ref{eqn5}) takes in this example. We have $U_1^{-1}HU_1= $

\begin{equation}
\left(
\begin{array}{cc}
\frac{i\mu_{3}^{*}B_{-}-i\mu_{3}B_{+}-B_3(1-|\mu_{3}|^2)}{2(1+|\mu_{3}|^2)} & i\mu_{3}B_3-\frac{1}{2}B_{-}-\frac{1}{2}\mu_{3}^2B_{+} \\
\frac{-2i\mu_{3}^{*}B_3-B_{+}-|\mu_{3}|^2B_{-}}{2(1+|\mu_{3}|^2)^2} & \frac{i\mu_{3}B_{+}-i\mu_{3}^{*}B_{-}+(1-|\mu_{3}|^2)B_3}{2(1+|\mu_{3}|^2)}
\end{array}
\right), 
\label{eqn15}
\end{equation}
and $iU_1^{-1}\dot{U}_1=$

\begin{equation}
\left(
\begin{array}{cc}
\frac{i\mu_{3}^{*}B_{-}+i\mu_{3}|\mu_{3}|^2B_{+}+2|\mu_{3}|^2B_3}{2(1+|\mu_{3}|^2)} &  i\mu_{3}B_3-\frac{1}{2}B_{-}-\frac{1}{2}\mu_{3}^2B_{+} \\
\frac{-2i\mu_{3}^{*}B_3-B_{+}-|\mu_{3}|^2B_{-}}{2(1+|\mu_{3}|^2)^2} & \frac{-\mu_{3}^{*}(i\mu_{3}^2B_{+}+iB_{-}+2\mu_{3}B_3)}{2(1+|\mu_{3}|^2)}
\end{array}
\right). 
\label{eqn16}
\end{equation}
Upon subtracting Eq.~(\ref{eqn16}) from Eq.~(\ref{eqn15}) to form $H_{\rm eff}$ in Eq.~(\ref{eqn5}), the off-diagonal terms cancel. The diagonal terms add to give precisely $(-B_3-i\mu_{3}B_{+}) \sigma_z /2$ to coincide with the last Eq.~(\ref{eqn8}) which gives the phase $\mu_1$. Our analysis for the triad choice $(J_{+}, J_{-}, J_3)$ in Eq.~(\ref{eqn6}) could also have been carried out for other choices of three linearly independent operators, for example, the Cartesian $(J_x, J_y, J_z)$ or an Euler-angle set. 

Having provided the spin-1/2 or SU(2) group's decomposition of the evolution operator into two factors $U_1$ and $U_2$ in detail as a pedagogical illustration, we note that we have given a closely parallel development for a more general SU($N$) for an arbitrary time-dependent Hamiltonian${}^{13}$. This construction reduces inductively the operator for $N$ to the one for $(N-1)$ with defining equations for complex parameters $z(t)$ which are analogs of $\mu_3$ and phases that are analogs of $\mu_1$. Further, while the $U(t)$ in Eq.~(\ref{eqn6}) is always unitary by construction, the individual $U_1$ and $U_2$ above are not but can also be unitarized. This is accomplished above by separating $\mu_1$ in $U_2$ in Eq.~(\ref{eqn14}) into real and imaginary parts, the latter as in Eq.~(\ref{eqn9}) incorporated into $U_1$ to make it unitary, leaving $U_2$, which depends on Re $\mu_1$, as a pure phase. A very similar construction${}^{13}$ for general SU($N$) also provides an explicitly unitary decomposition of the evolution operator, with $(N-1)$ phases, and again their decomposition into dynamical and geometric pieces as in Eq.~(\ref{eqn15}) and Eq.~(\ref{eqn16}).   

\section{Relation to variational principles and identities}

Finally, the connection of the effective Hamiltonian in Eq.~(\ref{eqn5}) to variational principles and their associated identities${}^{14}$ is also instructive. Following a general construction${}^{14}$, the time-dependent equation $i\dot{U}=HU$ can be converted into an identity

\begin{equation}
U(t) = U_t (t)-\int_0^t dt' L(t') [i\dot{U}_t (t')-H(t')U_t (t')],
\label{eqn17}
\end{equation}
where $U_t$ is a ``trial" function and $L$ a ``Lagrange adjoint" function \cite{VP} given by

\begin{equation}
L(t') = -iU(t)U^{-1}(t').
\label{eqn18}
\end{equation}
The identity in Eq.~(\ref{eqn17}) is easily verified upon doing an integration by parts and using Eq.~(\ref{eqn18}). Combining the two equations, we have

\begin{equation}
U(t)=U_t (t) +iU(t) \int_0^t dt' U^{-1} (t') [i\dot{U}_t (t')-H(t')U_t (t')],
\label{eqn19}
\end{equation}
and an associated variational principle for $U_{\rm var}$ which follows upon replacement of $U$ on the right-hand side of Eq.~(\ref{eqn19}) by its trial approximation $U_t$:

\begin{equation}
U_{\rm var} (t) = U_t (t) (1-i\int_0^t dt' U_t^{-1} (t')[-i\dot{U}_t (t') +H(t')U_t (t')]).
\label{eqn20}
\end{equation}
The right-hand side can be evaluated with any approximate solution $U_t$ and, as is clear from the derivation, will give a variationally correct approximation with only second order errors in $(U_t-U)$. Thus, $U_{\rm var}$ improves on $U_t$ which is only correct to first-order. Note the appearance of the $H_{\rm eff}$ of Eq.~(\ref{eqn5}) in the integrands of Eq.~(\ref{eqn19}) and Eq.~(\ref{eqn20}). 

This work has been supported by the National Science Foundation Grant 0243473 and by a Roy P. Daniels Professorship at LSU.

\end{document}